\documentclass[sigconf]{acmart}

\usepackage{xcolor} 
\usepackage{dblfloatfix}
\usepackage{tablefootnote}
\usepackage{threeparttable}

\usepackage{listings}
\usepackage[T1]{fontenc}

\usepackage{listings}
\usepackage{xcolor}
\usepackage{xcolor}

\lstdefinelanguage{text}{
    moredelim=**[is][\bfseries]{@}{@}, 
}

\usepackage{xspace}

\usepackage[most]{tcolorbox}
\tcbset{
  cikmquery/.style={
    enhanced,
    colback=gray!10,
    colframe=black!30,
    boxrule=0.3pt,
    arc=0pt,
    left=1mm,
    right=1mm,
    top=0.5mm,
    bottom=0.5mm,
    fonttitle=\bfseries\itshape,
    coltitle=black,
    title={Query},
    before skip=6pt,
    after skip=6pt,
    width=\columnwidth,    
    boxsep=1mm,
  }
}

\usepackage{mdframed}
\newmdenv[
  topline=false,
  bottomline=false,
  rightline=false,
  leftline=true,
  linecolor=gray,
  linewidth=2pt,
  backgroundcolor=gray!10,
  skipabove=6pt,
  skipbelow=6pt,
  innerleftmargin=6pt,
  innerrightmargin=6pt,
  innertopmargin=4pt,
  innerbottommargin=4pt,
]{querybox}

\usepackage{todonotes}
\AtBeginDocument{%
  }

\newcommand{\Method}{\textbf{\textit{InsertRank}}\xspace}
\setcopyright{acmlicensed}
\copyrightyear{2018}
\acmYear{2018}
\acmDOI{XXXXXXX.XXXXXXX}
\acmConference[InsertRank]{---}{Reasoning and Reranking}{Retrieval}
\acmISBN{978-1-4503-XXXX-X/2018/06}

\begin{document}


\title{\Method: LLMs can reason over BM25 scores to Improve Listwise Reranking}

\author{Rahul Seetharaman}
\affiliation{%
  \institution{UMass Amherst}
    \city{Amherst}
  \country{USA}
  }
\email{rseetharaman@umass.edu}

\author{Kaustubh D. Dhole}
\affiliation{%
  \institution{Emory University}
  \city{Atlanta}
  \country{USA}
}
\email{kdhole@emory.edu}

\author{Aman Bansal}
\affiliation{%
  \institution{UMass Amherst}
  \city{Amherst}
  \country{USA}
  }
\email{amanbansal@umass.edu}

\renewcommand{\shortauthors}{Seetharaman et al.}

\begin{abstract}

Large Language Models (LLMs) have demonstrated significant strides across various information retrieval tasks, particularly as rerankers, owing to their strong generalization and knowledge-transfer capabilities acquired from extensive pretraining. In parallel, the rise of LLM-based chat interfaces has raised user expectations, encouraging users to pose more complex queries that necessitate retrieval by ``reasoning'' over documents rather than through simple keyword matching or semantic similarity. While some recent efforts have exploited reasoning abilities of LLMs for reranking such queries, considerable potential for improvement remains. In that regards, we introduce \Method, an LLM-based reranker that leverages lexical signals like BM25 scores during reranking to further improve retrieval performance. \Method{} demonstrates improved retrieval effectiveness on -- BRIGHT, a reasoning benchmark spanning 12 diverse domains, and R2MED, a specialized medical reasoning retrieval benchmark spanning 8 different tasks. We conduct an exhaustive evaluation and several ablation studies and demonstrate that \Method{} consistently improves retrieval effectiveness across multiple families of LLMs, including GPT, Gemini, and Deepseek models. 
With Deepseek-R1, \Method{} achieves a score of 37.5 on the BRIGHT benchmark. and 51.1 on the R2MED benchmark, surpassing previous methods.
\end{abstract}




\maketitle

\section{Introduction}
Large language models (LLMs) have shown immense success across various tasks in natural processing and information retrieval. They have been successfully applied to reformulate queries~\cite{wang2023generative, dhole2024generative} and documents~\cite{nogueira2019document}, rerank documents~\cite{pradeep2023rankzephyr,10.1145/3626772.3657951} and products~\cite{productrank2025}, as well as train embeddings for dense indices~\cite{santhanam2022colbertv2}. They are pretrained on huge corpora, and often excel at a variety of search tasks, often without the need for additional fine-tuning.

In the context of reranking, they have been often used through 3 paradigms -- pointwise, pairwise, and listwise. The most common reranking models are cross encoders, which have a BERT backbone \cite{reimers2019sentencebertsentenceembeddingsusing}, ColBERT rerankers \cite{santhanam2022colbertv2}, which work on late interaction, and LLM-based rerankers. In many prior works, BM25 is utilized as a common first-stage retrieval system, from which the top-k documents are fed to the reranker. 

While there is prior work leveraging LLMs in all three settings (point, pair, and listwise), we focus on the listwise setting in this particular work. A listwise approach is advantageous as it consumes lesser number of tokens than the corresponding aggregated pointwise approaches, and allows the LLM to reason over all the documents at once. Moreover, with more and more LLMs having the ability to handle very long contexts, listwise reranking is becoming viable for LLM-based document reranking. In this work, we explore the use of LLMs on re-ranking passages in the context of complex reasoning-centric queries. To that end, we utilize two benchmarks - BRIGHT \cite{su2025brightrealisticchallengingbenchmark} and R2MED \cite{li2025r2medbenchmarkreasoningdrivenmedical} - a medical reasoning retrieval benchmark. Particularly, we experiment with the injection of BM25 scores into the prompt and demonstrate it as a useful signal for reasoning-centric LLM reranking. Injecting BM25 scores into the ranking input along with query and document has proven to be effective for BERT based cross encoders \cite{askari2023injectingbm25scoretext}. In this work, we demonstrate it is a useful signal to augment the reasoning capabilities of LLMs in the reranking setting.

In this work, we ask the following research question -- \textbf{\textit{Can incorporating lexical signals, such as retrieval scores, serve as effective clues for rerankers to improve retrieval effectiveness in reasoning tasks ?}}

Specifically, our work contributes the following:
\begin{enumerate}
\item  We introduce \Method{} -- a simple listwise reranking method that exploits BM25 retrieval scores to improve retrieval over reasoning queries
\item  We evaluate our method across multiple open and closed LLMs to demonstrate the effectiveness across two reasoning centric retrieval benchmarks - BRIGHT and R2MED
\item  We also conduct ablations and analyses along the following dimensions:
\begin{enumerate}
    \item Given the long context of the reranking inputs in the listwise setting, and tendencies of LLMs to favor context that is at the beginning and the end, we perform ablation experiments by shuffling the document order with and without BM25.
    \item We experiment with scale, normalization to examine their effects on the reasoning abilities of LLMs in the reranking context.
\end{enumerate}
\end{enumerate}

\textit{While many studies have focused on enhancing reasoning through fine-tuning and reinforcement learning methods, which rely heavily on labeled data, to the best of our knowledge, ours is the first work demonstrating improved retrieval effectiveness by integrating retrieval scores within a zero-shot setting using a listwise generative reranker.}

\section{Related work}
We now discuss related work to place our contributions in context.

\subsection{Retrieval for Complex Queries}


Reasoning-centric queries are often much more difficult and nuanced than those in traditional document retrieval, where keyword matching or semantic matching suffice. With LLMs becoming increasingly more powerful at reasoning and understanding, they become crucial for improving ranking and retrieval effectiveness for complex reasoning queries \cite{weller2025rank1testtimecomputereranking} \cite{su2025brightrealisticchallengingbenchmark}. BRIGHT \cite{su2025brightrealisticchallengingbenchmark} is a challenging benchmark of \~1300 queries across  11 domains and \~1M documents. Similarly, R2MED \cite{li2025r2medbenchmarkreasoningdrivenmedical} is a challenging benchmark of 876 queries across 8 tasks in the medical domain, which focuses on reasoning-centric retrieval. On BRIGHT, \citet{su2025brightrealisticchallengingbenchmark} have observed significant gains with query reformulation using GPT-4, Gemini, and other LLMs with a BM25 backbone.

There has also been growing research around training retrievers and rankers for reasoning-centric information retrieval. \citet{shao2025reasonirtrainingretrieversreasoning} finetune a LLama-8B model for complex reasoning queries. They also develop a synthetic data generation pipeline that produces complex hard negatives for finetuning a dense retrieval model. \citet{weller2025rank1testtimecomputereranking} leverage reasoning traces that are collected from Deepseek-R1 on the MSMARCO dataset and fine-tune small language models of varying sizes to achieve significant results on the BRIGHT benchmark. \cite{zhuang2025rankr1enhancingreasoningllmbased} leverage the GRPO technique to finetune language models of varying sizes (3B to 14B) for listwise reasoning centric reranking. \citet{yang2025rankktesttimereasoninglistwise} uses a listwise reranker finetuned on QwQ-32B and leverage a sliding window approach in a listwise setting to reduce the number of LLM calls compared to the pointwise setting. \citet{niu2024judgerankleveraginglargelanguage} leverages innovative prompting strategies with GPT-4 to score queries and documents in a pointwise setting.
While many works have focused on improving reasoning using finetuning and RL methods, to the best of our knowledge, ours is the first work to incorporate BM25 scores into the LLM prompt for a \textit{listwise zero shot setting.}

\subsection{LLM Based Reranking}
In the context of reranking, there are mainly three salient paradigms - pointwise, pairwise and listwise, with a fourth one namely setwise that has been recently introduced. 

\begin{itemize}
    \item \textbf{Pointwise}: Produce a score $s_j$ for each pair $(q_i,D_j)$ where $q_i$ is the ith query and $d_j$ is the jth document in the evaluation.
    \begin{equation}
\label{equ1}
    (q_i, D_j) \rightarrow \mathit{\textbf{M}} \rightarrow s_j 
\end{equation}
    \item \textbf{Pairwise}: Produce a preference score $s_j$ for each triple of the form
    $(q_i, D_j, D_k)$ ; the goal is to maximize the number of instances where $(q_i, D_j, D_k)>0$ when $D_j$ is more relevant than $D_k$ in the ground truth where $(q_i, D_j, D_k) > 0$ indicates $D_j$ is more relevant than $D_k$.
    \begin{equation}
\label{equ2}
(q, D_j, D_k) \rightarrow \mathit{\textbf{M}} \rightarrow s_j
\end{equation}
    \item \textbf{Listwise}: The goal in a listwise setting is to consider a query $q_i$ and a list of documents $D_1...D_n$ and produce a ranked list that takes in all the documents from the retriever at once. 
    \begin{equation}
\label{equ3}
(q, D_1, \ldots , D_n) \rightarrow \mathit{\textbf{M}} \rightarrow r_1 \, r_2 \, \ldots \, r_n
\end{equation}
\end{itemize}

where $r_1, r_2, ... r_n$ are the ranked list of documents or their identifiers.


\subsection{Leveraging numerical information in language models}
There have been several prior works studying how both BERT based (encoder) and decoder only LLMs understand numerics. \cite{liu2025mathematicallanguagemodelssurvey} provide a comprehensive survey of mathematical LLMs, covering CoT, tool use, instruction tuning, etc. Similarly, \cite{ahn-etal-2024-large} provides an overview of LLM abilities in problem solving, math reasoning, geometry and so on. \citet{askari2023injectingbm25scoretext} fine-tuned a BERT based cross encoder reranker by injecting BM25 scores along with the document tokens and found improvements over a pointwise cross encoder setup. However, it is unclear whether BM25 scores can boost recent LLM capabilities under more realistic settings -- namely without fine-tuning, and in listwise reranking. Our work demonstrates the \textbf{effectiveness of injecting BM25 scores in a zero shot listwise setting with no finetuning}. 
\begin{table*}[!htbp]

\centering
\caption{Performance in BRIGHT benchmark (P - Pointwise, L - Listwise, only retrieval if neither P nor L mentioned)}
\begin{tabular}{lccccccccccccc}
\toprule
 & \textbf{Bio} & \textbf{Earth} & \textbf{Econ} & \textbf{Psy} & \textbf{Robot} & \textbf{Stack} & \textbf{Sust} & \textbf{LC} & \textbf{Pony} & \textbf{AoPS} & \textbf{TheoQ} & \textbf{TheoT} & \textbf{Avg} \\
\midrule
BM25 & .192 & .271 & .149 & .125 & .135 & .165 & .152 & .244 & .079 & .060 & .130 & .069 & .148 \\
BM25 on GPT-4o CoT & .536 & .536 & .243 & .386 & .188 & .227 & .259 & .193 & .177 & .039 & .189 & .202 & .265 \\
Rank-R1-14B GRPO \textsuperscript{\$} & .312 & .385 & .212 & .264 & .226 & .189 & .275 & .092 & .202 & .097 & .119 & .092 & .205 \\
ReasonIR reranker(P) \textsuperscript{\$} & .582 & .532 & .320 & .436 & .288 & .376 & .360 & .332 & .348 & .079 & .326 & .450 & .369 \\

JudgeRank Ensemble(P) \textsuperscript{\$} & .607 & .587 & .354 & .476 & .282 & .297 & .419 & .202 & .327 & .086 & .259 & .362 & .355 \\
Rank1-32B (P) \textsuperscript{\$} & .497 & .358 & .220 & .375 & .225 & .217 & .350 & .188 & .325 & .108 & .229 & .437 & .294 \\
Rank-K (Sliding Window) (L) \textsuperscript{\$} & .506 & .442 & .353 & .454 & .247 & .307 & .384 & .327 & .242 & .090 & .383 & .281 & .335 \\

\hline
Gemini 2.0 flash (L) & .556 & .520 & .286 & .488 & .300 & .315 & .415 & .203 & .303 & .052 & .225 & .348 & .334 \\
\textbf{+\Method{}}  & \textbf{.594} & \textbf{.556} & \textbf{.309} & .512 & .307 & .305 & .391 & \textbf{.224} & .265 & \textbf{.070} & \textbf{.240} & \textbf{.371} & \textbf{.345} \\
\hline
Gemini 2.5 flash (L) & .540 & .543 & .254 & .403 & .231 & .306 & .302 & .224 & .195 & .053 & .224 & .241 & .293 \\
\textbf{+\Method{}}  & \textbf{.596} & \textbf{.564} & \textbf{.298} & \textbf{.496} & \textbf{.298} & \textbf{.326} & \textbf{.391} & .207 & \textbf{.253} & .053 & .237 & \textbf{.377} & \textbf{.341} \\
\hline
GPT-4o (L) & .598 & .556 & .299 & .530 & .346 & .328 & .426 & .229 & .311 & .077 & .308 & .409 & .368 \\
\textbf{+\Method{}}  & \textbf{.626} & \textbf{.578} & \textbf{.326} & .505 & .317 & \textbf{.349} & .422 & \textbf{.264} & .226 & \textbf{.091} & \textbf{.328} & \textbf{.426} & \textbf{.371} \\
\hline
Deepseek-r1 (L) & .530 & .507 & .325 & \textbf{.530} & .349 & \textbf{.365} & \textbf{.472} & \textbf{.224} & \textbf{.243} & .089 & .343 & .487 & .372 \\
\textbf{+\Method{}}  & \textbf{.582} & \textbf{.527} & \textbf{.347} & .517 & \textbf{.362} & .348 & .428 & .220 & .229 & \textbf{.102} & \textbf{.351} & \textbf{.489} & \textbf{.375} \\
\bottomrule
\end{tabular}

\begin{tablenotes}
\footnotesize
\item \$ - for baseline comparisons, we have taken the best results from each of the above works
\end{tablenotes}

\label{tab:bright_results}

\end{table*}

\begin{table*}[!htbp]
\centering
\caption{Performance in R2MED benchmark}
\begin{tabular}{lccccccccc}
\toprule
 & \textbf{Bioinfo} & \textbf{Biology} & \textbf{IIYi-Clin} & \textbf{Med-Sci} & \textbf{MedQA-Diag} & \textbf{MedXQA} & \textbf{PMC-Clin} & \textbf{PMC-Treat} & \textbf{Avg} \\
\midrule
\hline
BM25 (HyDE) \textsuperscript{\$} & .430 & .573 & .104 & .412 & .496 & .265 & .328 & .406 & .377 \\
NVEmbed + GPT-4o CoT \textsuperscript{\$} & .336 & .542 & .508 & .231 & .361 & .474 & .485 & .213 & .394 \\
NVEmbed + o3-mini CoT \textsuperscript{\$}  & .340 & .559 & .513 & .290 & .403 & .490 & .509 & .205 & .414 \\
\hline
\hline
Gemini 2.0 flash (L) & \textbf{.588} & .534 & \textbf{.156} & .559 & \textbf{.566} & \textbf{.371} & \textbf{.477} & .593 & .480 \\
\textbf{+\Method{}} & .581 & \textbf{.603} & .145 & \textbf{.566} & .562 & .351 & .450 & \textbf{.614} & \textbf{.484} \\
\hline

Gemini 2.5 flash (L) & .660 & .577 & \textbf{.220} & .580 & .578 & .427 & \textbf{.535} & .624 & .525 \\
\textbf{+\Method{}}  & \textbf{.684} & \textbf{.630} & .185 & \textbf{.587} & \textbf{.589} & \textbf{.467} & .524 & \textbf{.628} & \textbf{.537} \\
\hline
GPT-4o (L) & \textbf{.627} & .567 & .186 & \textbf{.594} & .594 & .377 & .423 & \textbf{.637} & .501 \\
\textbf{+\Method{}}  & .622 & \textbf{.602} & .186 & .590 & \textbf{.596} & \textbf{.382} & \textbf{.429} & .626 & \textbf{.504} \\
\hline
Deepseek-r1 (L) & .615 & .559 & \textbf{.201} & \textbf{.612} & .581 & \textbf{.394} & .437 & \textbf{.664} & .508 \\
\textbf{+\Method{}}  & \textbf{.639} & \textbf{.576} & .196 & .600 & \textbf{.605} & .376 & \textbf{.442} & .652 & \textbf{.511} \\
\bottomrule
\end{tabular}
\label{tab:r2med_results}
\begin{tablenotes}
\footnotesize
\item \$ - for baseline comparisons, we have taken the best results from each of the above works
\end{tablenotes}

\end{table*}

\section{Proposed Method}
We now describe our proposed method.
\Method{} involves injecting the retriever's BM25 score into the listwise reranking setting.
    \begin{equation}
\label{equ3}
(q, D_1, b_1 \ldots , D_n, b_n) \rightarrow \mathit{\textbf{M}} \rightarrow r_1, r_2 \ldots r_l
\end{equation}

Here, $q$ is the query, $D_1, D_2, ... D_n$ is the list of documents passed to the reranker, $b_1, b_2, ... b_n$ is the BM25 scores associated with each document, and $r_1, r_2, ... r_n$ is the reranked list of document identifiers. 

In addition, in our experiments with BRIGHT and R2MED, the documents are passed in decreasing order of their BM25 scores, i.e $b_1 > b_2 > ... b_n$

For the BRIGHT benchmark, inspired by the Rank-1 paper, we leverage their queries augmented by GPT-4 chain of thought (CoT) as it is reported to give the best NDCG@10 scores on BM25. Similarly, for the R2MED benchmark, we leverage the HyDE query reformulation \cite{gao2022precisezeroshotdenseretrieval} mentioned in their work, where a hypothetical document is leveraged as a query reformulator. as it is reported to give the best NDCG@10 scores in their BM25 first stage retrieval setting. In a listwise setting with BM25 score injection, the queries and documents along with the scores are passed as follows,

\begin{querybox}
<instructions> You are also given the BM25 scores from a lexical retrieval sytem.
<query> <{doc\_1, BM25 score: s\_1}, {doc\_2, BM25 score: s\_2}, .... {doc\_n, BM25 score: b\_n}>
\end{querybox}

Other than the ablation setting in Table \ref{tab:r2med_shuffle}, \textit{documents are ordered by decreasing order of BM25 scores}. 

In a reasoning setting, LLMs are often known to have issues like hallucinations \cite{hallucinationssurvey}, incorrect reasoning \cite{lee2024llmhallucinationreasoningzeroshot}, brittleness with respect to changing numbers and names \cite{mirzadeh2024gsmsymbolicunderstandinglimitationsmathematical}. Overthinking is also established as a common issue in reasoning models - a tendency where LLMs tend to produce very verbose reasoning chains for simpler problems leading to issues like concept drift \cite{bao2025learningstopoverthinkingtest} \cite{chen2025think23overthinkingo1like}. By providing a critical lexical relevance signal like BM25 scores, the goal is to ground the reasoning and prevent the model from running into issues like overthinking and concept drift and ground the reasoning process with respect to the first stage retriever.

By injecting the BM25 scores in the LLM reranking step, we provide a low cost solution for improving reasoning centric LLM reranking. Our solution produces consistent improvements across two reasoning centric retrieval benchmarks with no extra cost of finetuning and negligible additional token costs.

For all our experiments, we utilize the official repository of BRIGHT and R2MED. For the LLM implementations of Gemini-2.0-flash, GPT-4o and Deepseek-R1, we use their respective official APIs.

We compare \Method{} with multiple baselines:
\begin{enumerate}
    \item ReasonIR \cite{shao2025reasonirtrainingretrieversreasoning}, which trains a retriever with hard synthetic negatives and incorporates a hybrid BM25 and pointwise reranking setup on top of it
    \item Rank1-32B \cite{weller2025rank1testtimecomputereranking}, which finetunes a 32B parameter model from Deepseek-R1's reasoning traces 
    \item Rank-R1 GRPO \cite{zhuang2025rankr1enhancingreasoningllmbased}, which finetunes an  LLM using reinforcement learning based methods for listwise ranking
    \item JudgeRank \cite{niu2024judgerankleveraginglargelanguage}, which is a prompting approach for pointwise reranking.
    \item Rank-K~\cite{yang2025rankktesttimereasoninglistwise} which finetunes a QwQ-32B model and uses a sliding window style listwise ranking approach
\end{enumerate}

\section{Results}
The results of our experiment are as shown in Table \ref{tab:bright_results} and \ref{tab:r2med_results}. The top performing setting scores an average of \textbf{37.5} on the BRIGHT benchmark. We observe consistent gains by injecting BM25 scores into the prompt on multiple LLM families - Gemini, GPT-4 and Deepseek. The results show consistent improvement over a vanilla LLM listwise reranking with just queries and documents. By injecting BM25 scores into the prompt, we show gains of 3.2\% on Gemini 2.0 flash, 16.3\% on Gemini 2.5 flash, 0.8\% on GPT-4o and Deepseek-r1 compared to just using the raw queries and documents. While we are able to use full length of the documents in the Gemini models, due to context length limitations, we use only the first 1800 tokens for the GPT-4o and Deepseek series of models.

We observe similar gains in the R2MED benchmark, with BM25 injection consistently surpassing the ranking quality on average compared to the vanilla listwise setting which takes just the documents into the prompt. The gains demonstrated are 0.8\% for Gemini 2.0 flash, 2.2\% in Gemini 2.5 flash and 0.5\% gains in GPT-4o and Deepseek family of models.

\section{Ablations}

In this section, we analyze the effectiveness of our proposed method in the context of 1) normalization 2) shuffling input documents
\subsection{Scaling and normalization of BM25}
We additionally also examine the effect of different scales and how LLMs perceive them. Since BM25 scores are normally not restricted to a particular range, we perform an experiment with normalized BM25 scores. Similar to the previous experiments, we do two settings -- one with BM25 injected and one without. Finally, we also examine the effect of scaling wherein normalized scores are scaled from 0-100. 

The ablation results are reported for both R2MED and BRIGHT on the Gemini 2.0 flash model. As evidenced in \ref{tab:r2med_norm}, R2MED shows around 0.4\% decrease when BM25 scores are normalized from 0-1 and 0.8\% increase when BM25 scores are normalized from 0-100. While there is a small decrease in the 0-1 normalization setting, the 0-100 normalization, shows a marginal improvement.

Similarly for BRIGHT, we observe similar marginal performance gains when using a 0-100 normalization and very slight decrease when scores are normalized from 0-1. The results listed in table \ref{tab:r2med_norm} indicate a 0.58\% decrease when normalizing from 0-1 and 0.5\% increase when normalizing using 0-100. 
With NDCG@10 scores across normalization also beating the vanilla listwise setting , \textbf{the results demonstrate the robustness of \Method{} to normalization and scaling.}

\begin{table}[]
\centering
\caption{Effect of normalized BM25 scores}
\begin{tabular}{|l|l|c|}
\hline
&  \textbf{BRIGHT} & \textbf{R2MED} \\
\hline
 \textbf{Setting}& \textbf{Avg}&\textbf{Avg}\\\hline
\hline
Vanilla (no scores) & .334 & .480 \\ 
\hline
Raw BM25 scores & \textbf{.345} & .484 \\ \hline
0-1 scale &  .340 &.482 \\
\hline
0-100 scale &  .342 &\textbf{.488} \\
\hline
\end{tabular}
\label{tab:r2med_norm}
\end{table}

\subsection{Shuffling order of documents}
LLMs are well known to prefer documents at the beginning and end of context \cite{liu2023lostmiddlelanguagemodels}. In order to validate the effectiveness of the proposed approach we shuffle the document tuples, where each tuple is of the form ${D,B}$ where $D$ is the document and $B$ is the BM25 score associated with the document. Similar to the previous experiments, we perform two settings, one with the BM25 scores injected and one without it. Unlike, normalization, we observe divergent results - while BRIGHT benchmark demonstrates robustness in the BM25 injection and shows gains, R2MED on the other hand, shows a consistent decrease when the documents are shuffled.  Similar to the previous ablation, we report results for both BRIGHT and R2MED on Gemini 2.0 flash. As evidenced in \ref{tab:r2med_shuffle}, BRIGHT in the shuffled setting with BM25 injection demonstrates a 9.4\% increase relative to the vanilla setting. However, there is a 1.1 points absolute decrease compared to the original setting, where documents are passed in decreasing order of BM25 scores. This establishes that listwise reranking methods in general are very sensitive to initial ordering for reasoning centric retrieval/reranking.
\begin{table}[H]
\centering
\caption{Effect of shuffling on R2MED}
\begin{tabular}{|l|c|l|}\hline
 & \textbf{BRIGHT} & \textbf{R2MED} \\\hline
\hline
\textbf{Setting} & \textbf{Avg}  &\textbf{Avg} \\
\hline
Vanilla (ordered w/o BM25) & \textbf{.334} & .480  \\ 
\hline
Shuffled & .285 & \textbf{.488}\\
\hline
Shuffled w/ BM25  &.322 & .445  \\
\hline
\end{tabular}
\label{tab:r2med_shuffle}
\end{table}

\section{Conclusion and Future Work}
In this paper, we present \Method{}, a simple test time strategy that incorporates feedback from the first stage retriever, improving LLM reranking for reasoning centric retrieval tasks. By incorporating BM25 based lexical signals, we are able to consistently improve retrieval effectiveness across a wide variety of tasks on multiple reasoning centric retrieval benchmarks.

Our work also opens up new avenues for research as to what other low-cost relevance signals and additional meta-data could reasoning behavior take advantage from to improve reranking further? LLMs' reasoning abilities at inference time enable improved document ranking performance by supporting zero-shot reasoning over documents with their BM25 scores, a phenomenon not previously observed in non-reasoning zero-shot scenarios. This indicates that the reasoning capacity of LLMs offers a promising avenue for exploiting possibly other metadata, potentially enriching document representation and retrieval.

Moreover, our work also encourages other reranking applications to look for additional metadata, such as in distillation, which depend on exploiting reasoning traces of larger models.

\newpage

\bibliographystyle{ACM-Reference-Format}
\bibliography{sample-sigconf}

\section{Appendix}
\label{appendix}

\subsection{Effect of shuffling}

Tables \ref{tab:r2med_shuffle_full} and \ref{tab:bright_shuffle_full} illustrate the performance differences while passing in a shuffled order of the documents and their associated scores.

\begin{table*}[!htbp]
\centering
\caption{Effect of shuffling on R2MED}
\begin{tabular}{lccccccccc}
\toprule
Setting & \textbf{Bioinfo} & \textbf{Biology} & \textbf{IIYi-Clin} & \textbf{Med-Sci} & \textbf{MedQA-Diag} & \textbf{MedXQA} & \textbf{PMC-Clin} & \textbf{PMC-Treat} & \textbf{Avg} \\
\midrule
Shuffled & \textbf{.621} & \textbf{.533} & \textbf{.150} & \textbf{.569} & \textbf{.564} & \textbf{.369} & \textbf{.491} & \textbf{.610} & \textbf{.488} \\
Shuffled w/ BM25 & .586 & .518 & .084 & .533 & .505 & .313 & .451 & .568 & .445 \\
\bottomrule
\end{tabular}
\label{tab:r2med_shuffle_full}
\end{table*}

\begin{table*}[!htbp]
\centering
\caption{Effect of shuffling on BRIGHT}
\begin{tabular}{lccccccccccccc}
\toprule
Setting & \textbf{Biology} & \textbf{Earth Sci} & \textbf{Econ} & \textbf{Psych} & \textbf{Robot} & \textbf{Stack} & \textbf{Sust} & \textbf{LC} & \textbf{Pony} & \textbf{AoPS} & \textbf{TheoQ} & \textbf{TheoT} & \textbf{Avg} \\
\midrule
Shuffled & .438 & .380 & .254 & .449 & .254 & .220 & .341 & .210 & \textbf{.281} & .043 & .220 & .334 & .285 \\
Shuffled w/ BM25 & \textbf{.508} & \textbf{.444} & \textbf{.270} & \textbf{.489} & \textbf{.296} & \textbf{.285} & \textbf{.356} & \textbf{.255} & .225 & \textbf{.077} & \textbf{.264} & \textbf{.394} & \textbf{.322} \\

\bottomrule
\end{tabular}
\label{tab:bright_shuffle_full}
\end{table*}

\subsection{Effect of normalization}

Tables \ref{tab:bright_norm_full} and \ref{tab:r2med_norm_full} show the effect of normalization on all splits.

\begin{table*}[!htbp]
\centering
\caption{Performance of BM25+Norm with different scaling factors across domains on Gemini benchmark}
\begin{tabular}{lccccccccccccc}
\toprule
\textbf{Model} & \textbf{Bio} & \textbf{Earth} & \textbf{Econ} & \textbf{Psy} & \textbf{Robot} & \textbf{Stack} & \textbf{Sust} & \textbf{LC} & \textbf{Pony} & \textbf{AoPS} & \textbf{TheoQ} & \textbf{TheoT} & \textbf{Avg} \\
\midrule
0-1 scale & .531 & .454 & .302 & .504 & .291 & .363 & .333 & .250 & .314 & .069 & .276 & .394 & .340 \\
0-100 scale & .541 & .444 & .298 & .518 & .296 & .346 & .356 & .253 & .300 & .077 & .282 & .394 & \textbf{.342} \\
\bottomrule
\end{tabular}
\label{tab:bright_norm_full}
\end{table*}

\begin{table*}[!htbp]
\centering
\caption{R2MED normalized BM25 scores}
\begin{tabular}{lccccccccc}
\toprule
Setting & \textbf{Bioinfo} & \textbf{Biology} & \textbf{IIYi-Clin} & \textbf{Med-Sci} & \textbf{MedQA-Diag} & \textbf{MedXQA} & \textbf{PMC-Clin} & \textbf{PMC-Treat} & \textbf{Avg} \\
\midrule
0-1 scale & .588 & .573 & .130 & .560 & .576 & .351 & .465 & .614 & .482 \\
0-100 scale & \textbf{.621} & .533 & \textbf{.150} & \textbf{.569} & .564 & \textbf{.369} & \textbf{.491} & .610 & \textbf{.488} \\
\bottomrule
\end{tabular}
\label{tab:r2med_norm_full}
\end{table*}

\subsection{Reasoning Traces}

We also provide examples of reasoning traces that demonstrate how an LLM (Deepseek-R1) leverages and thinks with BM25 reasoning. 

\subsubsection{Query}
\begin{querybox}
How to count reads in bam per bed interval with  bedtools/nI recently installed Ubuntu 16.04 (because I was still using 12.04). But it seems my bedtools scripts don't work properly anymore. I can't figure out how to use the new bedtools for my old ways. What I want to do is get the number of reads from a bam file, per interval from a bed file. It was very simple with my old version of Ubuntu and bedtools:bedtools coverage -abam file.bam -b All\_peaks.bed > file.cov.txt.But this doesn't seem to work anymore. I used to get small files as results (3MB), but now huge files are created (4GB) including read names?? I am using bedtools v2.25 now., 
\end{querybox}

\subsubsection{With BM25 reasoning (on R2MED)}

\begin{querybox}
    ...

Looking at the provided passages, I need to find which ones discuss changes in bedtools coverage behavior...

...

Looking through the provided documents, Document 2's content about the coverage tool's change in behavior might be relevant.

...

Looking back at the provided passages, Document 8 mentions the -split option for coverage calculations. Document 21 talks about the -bga option for genomecov to report all positions, which is more verbose. But the user's problem is with coverage, not genomecov.

...

Looking at Document 1 again: it's about genomecov's options, not coverage.

...

\textbf{Let's look at the highest BM25 documents again. Document 2 talks about coverage tool changes, which is relevant}. Document 8 discusses coverage options including -split and -strand. Document 1 is about genomecov.

...

Given the information in the provided passages, the most relevant documents are:

Document 2: Discusses changes in the coverage tool's behavior in version 2.24.0, including processing order and -sorted option.

Document 8: Mentions options for coverage calculation, including -split...

Document 21: Talks about genomecov's -bga option...

...

Given the confusion, the most likely relevant passage is Document 2, which discusses changes in the coverage tool's behavior.

...

\textbf{However, considering the BM25 scores, Document 2 has the second highest score and discusses coverage tool changes, which is likely relevant.} Document 8 discusses coverage options and could be related. Document 21 talks about genomecov's -bga option...

In conclusion, the most relevant passages are likely Document 2 and Document 8.
\end{querybox}

\subsubsection{Without BM25 reasoning on R2MED}

\begin{querybox}

The user is using:
bedtools coverage -abam file.bam -b Allpeaks.bed > file.cov.txt
Previously, this produced a 3MB summary file, but now it generates a 4GB file including read names.

...

Document 2 notes a critical change in bedtools version 2.24.0:

"coverage is now computed for the intervals in the A file based on overlaps with the B file, rather than vice versa."

That’s key.

In the old version, -abam file.bam -b peaks.bed treated the BED intervals as targets for coverage. In the new version, bedtools computes coverage for A (the BAM) over B (the BED). Since BAM entries are individual reads, the tool now outputs each read along with overlap details — hence the explosion in file size and appearance of read names.

...

The correct fix in v2.25 is:

bedtools coverage -a All\_peaks.bed -b file.bam > file.cov.txt
...

This change in semantics (A vs. B) is the root cause, and is well-documented in Document 2. Other documents (8, 14, 21) mention flags like -split, -d, or -bga but aren’t directly relevant here.

\end{querybox}

\subsubsection{Gold Answer}

\begin{querybox}
    Uppers" and "downers" are colloquial terms for drugs with stimulant and depressant effects, respectively.
...
The primary concern in overdose is overstimulation or suppression of cardiovascular/respiratory systems.
...
One combination is a "speedball" (cocaine + heroin/morphine). The mix gives euphoria and masks side effects, so users may take higher doses.
...
However, drug durations differ: when the stimulant wears off, opioid effects persist, leading to unexpected respiratory depression and death.
...
Alcohol can worsen effects due to paradoxical reactions; alcohol + cocaine is also linked to overdose deaths.
...
Stimulant + depressant combos mask warning signs and increase overdose and abuse potential.
...
\end{querybox}

\subsection{Full Prompts}

\subsubsection{Full prompts for BRIGHT}

\begin{querybox}
The following passages are related to query: {cur\_query}

\{retriever\_prompt\}\{doc\_string\}
First identify the essential problem in the query.
Think step by step to reason about why each document is relevant or irrelevant.
Rank these passages based on their relevance to the query.
Please output the ranking result of passages as a list, where the first element is the id of the most relevant passage, the second element is the id of the second most element, etc.
Please strictly follow the format to output a list of \{topk\} ids corresponding to the most relevant \{topk\} passages:
```json
[...]
\end{querybox}

\subsubsection{Full prompts for R2MED}

\begin{querybox}
The following passages are related to the query: "{cur\_query}".

\{retriever\_prompt\}

\{doc\_string\}
First, identify the essential problem or topic in the query.
Think step by step to reason about why each document is relevant or irrelevant..
Rank these passages based on their relevance to the query.
Please output the ranking result of passages as a list, where the first element is the id of the most relevant passage, the second element is the id of the second most element, etc.
Finally, output a ranked list of the top {topk} most relevant passages by their index number.
Please strictly follow the format to output a list of \{topk\} ids corresponding to the most relevant \{topk\} passages:
```json
[...]
\end{querybox}

The retriever prompt which introduces BM25 is defined as follows,

\begin{querybox}

You are also given the BM25 scores from a lexical retriever for each document.

[1]. \{doc\_text\_1\} BM25 score: \{score\_1\} 

[2]. \{doc\_text\_2\} BM25 score: \{score\_2\} 

[3]. \{doc\_text\_3\} BM25 score: \{score\_3\} 

...

First identify the essential problem in the query.
Think step by step to reason about why each document is relevant or irrelevant.
Rank these passages based on their relevance to the query.
Please output the ranking result of passages as a list, where the first element is the id of the most relevant passage, the second element is the id of the second most element, etc.
Please strictly follow the format to output a list of {topk} ids corresponding to the most relevant {topk} passages:
```json
[...]

\end{querybox}

\end{document}